\documentclass[prb,twocolumn,aps,10pt,eqsecnum,superscriptaddress]{revtex4-1}

\usepackage{amsmath}
\usepackage{graphicx}
\usepackage{rotating}
\usepackage{multirow}
\usepackage{latexsym}
\usepackage{textcomp}
\usepackage{verbatim}
\usepackage{color}
\usepackage{pict2e}
\usepackage{bm}
\usepackage{subfigure}
\usepackage{everysel}
\usepackage{keyval}
\usepackage{ragged2e}
\usepackage{dsfont}
\usepackage{amssymb}
\usepackage{enumitem}
\usepackage{pstricks}
\usepackage{mathtools}
\usepackage{hyperref}

\newcommand{\beq}{\begin{equation}} 
\newcommand{\tr}{\text{tr}}
\newcommand{\eeq}[1]{\label{#1} \end{equation}}

\newcommand\bmb{\left( \begin{matrix}}
\newcommand\emb{\end{matrix} \right)}

\renewcommand{\[}{\left[}
\renewcommand{\]}{\right]}
\renewcommand{\Im}{\mathrm{Im}}

\def\LSCO{La$_{2-x}$Sr$_x$CuO$_4$}
\def\LBCO{La$_{2-x}$Ba$_x$CuO$_4$}
\def\YBCO{YBa$_2$Cu$_3$O$_{6+x}$}

\def\BSCCO{Bi$_2$Sr$_2$CaCu$_2$O$_{8+\delta}$}

\def\C60{A$_x$C$_{60}$}

\def\HgCu3{HgCa$_2$Cu$_3$O$_{8+y}$}
\def\HgCu4{HgBa$_2$Ca$_3$Cu$_4$O$_{10+y}$}
\def\TlCu{Tl$_2$Ba$_2$CuO$_{6+\delta}$}
\def\TlCu3{Tl$_2$Ba$_2$Ca$_2$Cu$_3$O$_{10+y}$}
\def\TlCu4{Tl$_2$Ba$_2$Ca$_3$Cu$_4$O$_{12+y}$}

\def\BiCu3{Bi$_2$Sr$_2$Ca$_{2}$Cu$_3$O$_y$}

\def\8LSCO{La$_{1.88}$Sr$_{.12}$CuO$_4$}
\def\110LNSCO{La$_{1.5}$Nd$_{0.4}$Sr$_{0.1}$CuO$_{4}$}
\def\stage4LCO{La$_{2}$CuO$_{4+\delta}$}
\def\Y248{YBa$_2$Cu$_4$O$_8$}

\def\NbSe2{NbSe$_2$}
\def\TaSe2{TaSe$_2$}
\def\TiSe2{TiSe$_2$}
\def\NaCoOH2O{Na$_{0.3}$CoO$_{2y}$H$_2$O}
\def\MgB2{MgB${}_2$}

\def\URu2Si2{URu$_2$Si$_2$}

\def\Ba122{Ba(Fe$_{1-x}$Co$_x$)$_2$As$_2$}

\begin{document}

\title{Higgs Modes in the Pair Density Wave Superconducting State}
\author{Rodrigo Soto-Garrido }
\affiliation{Facultad de Ingenier\'ia y Tecnolog\'ia, Universidad San Sebasti\'an, Bellavista 7, Santiago 8420524, Chile} 
\author{Yuxuan Wang}
\affiliation{Department of Physics and Institute for Condensed Matter Theory, University of Illinois at Urbana-Champaign, 1110 
West Green Street, Urbana, IL  61801-3080, USA} 
\author{Eduardo Fradkin}         
\affiliation{Department of Physics and Institute for Condensed Matter Theory, University of Illinois at Urbana-Champaign, 1110 
West Green Street, Urbana, IL  61801-3080, USA} 
\author{S. Lance Cooper}
\affiliation{Department of Physics and Frederick Seitz Materials Research Laboratory, University of Illinois at Urbana-Champaign, 1110 
West Green Street, Urbana, IL  61801-3080, USA}

\date{\today}

\begin{abstract}
The pair density wave (PDW) superconducting state has been proposed to explain the layer-decoupling effect observed in the La$_{2-x}$Ba$_x$CuO$_4$ compound at $x=1/8$ (Phys. Rev. Lett. 99, 127003). In this state the superconducting order parameter is spatially modulated, in contrast with the usual superconducting (SC) state where the order parameter is uniform. In this work, we study the properties of the amplitude (Higgs) modes in a 
unidirectional PDW state. 
To this end we consider a phenomenological model of PDW  type states  coupled to a Fermi surface of fermionic quasiparticles. In contrast to conventional superconductors that have a single Higgs mode, unidirectional PDW superconductors have two Higgs modes.
While in the PDW state the Fermi surface largely remains gapless, we find that the damping of the PDW Higgs modes into fermionic quasiparticles requires exceeding an energy threshold. We show that this suppression of damping in the PDW state is due to kinematics. As a result, only one of the two Higgs modes is significantly damped.
In addition, motivated by the experimental phase diagram, we discuss the mixing of Higgs modes in the coexistence regime of the PDW and uniform SC states.
These results should be observable directly in a Raman spectroscopy, in momentum resolved electron energy loss spectroscopy, and in resonant inelastic X-ray scattering, thus providing evidence of the PDW states.
\end{abstract}

\maketitle

\section{Introduction}

In the conventional theory of superconductivity by Bardeen-Cooper-Schrieffer (BCS), \cite{Bardeen-1957} the Cooper pairs have a zero center of mass momentum and  the superconducting (SC) order parameter is uniform in space.
In this paper we will examine the physical properties of pair-density-wave (PDW) SC states, SC states whose order parameters have finite momentum, which generically are non-uniform. 
{PDW states are phases in which superconducting and charge-density-wave (CDW) and/or spin-density-wave (SDW) orders are intertwined.\cite{Berg-2007,Berg-2009,fradkin-2014}}
PDW SC states have been proposed\cite{Berg-2007} to explain many experimental features of a family of cuprate high-temperature superconductors (HTSC), notably {\LBCO} near doping $x=1/8$ and the observed dynamical layer decoupling of  transport properties\cite{Li-2007,Tranquada-2008} (both at zero and with finite magnetic fields), and the {\LSCO} materials in the underdoped regime at large enough magnetic fields.\cite{abbamonte-2005,lake-2002,Schafgans-2010,Schafgans-2010b} 
Evidence for PDW-type SC states has also been found in the heavy-fermion material CeRhIn$_5$ at high magnetic fields.\cite{Park-2012}  
Charge and spin stripe phases, both static or ``fluctuating'', have been observed experimentally in HTSC materials for quite some time,\cite{kivelson-2003} including the more recently observed CDWs  in {\YBCO} \cite{Wu-2011,chang-2012,Ghiringhelli-2012,leboeuf-2012,Wu-2013,blackburn-2013,Blanco-Canosa-2014,Comin-2015,Achkar-2016,Jang-2016}, in {\BSCCO},\cite{lawler-2010,silva_neto-2012,Silva_Neto-2014,Hamidian-2015,Vig-2015,Mesaros-2016}, and in HgBa$_2$CuO$_{4+\delta}$.~\cite{Greven-2014}
  Whether PDW type phases arise in HTSC other than the lanthanum materials is presently not known. However, the recent observation of a modulation of the superfluid density deep in the $d$-wave SC phase of  {\BSCCO}   by  scanned Josephson tunneling  microscopy \cite{Hamidian-2016} supports that possibility. The purpose of this paper is to establish theoretically spectroscopic properties of PDW phases that (hopefully) will help to identify these states.

Non-uniform SC states have a position-dependent complex spin-singlet order parameter field $\Delta(\bm r)$ of the form:
\begin{equation}
\Delta(\bm r)=\Delta_0(\bm r) +\Delta_{\bm Q}(\bm r) e^{i {\bm Q} \cdot {\bm r}}+\Delta_{-\bm Q}(\bm r) e^{-i {\bm Q} \cdot {\bm r}}+\ldots
\label{eq:expansion}
\end{equation}
For simplicity, here we consider only states with unidirectional order with ordering wave vector ${\bm Q}$. The generalization to more complex types of ordered patterns is straightforward. 
SC states with an order parameter of the form of Eq.\eqref{eq:expansion} were first
considered long ago by 
Fulde and Ferrell \cite{Fulde-1964} (FF) and Larkin and Ovchinnikov \cite{Larkin-1964} (LO). These authors showed that 
in a BCS model with a  Zeeman coupling to a  uniform external magnetic field it is possible to have a SC state with  Cooper pairs condensing with a finite center of mass momentum.

Here we will focus  on SC states with a finite ordering wave vector $\bm Q$ of the form of Eq.\eqref{eq:expansion}, arising {\em in the absence} of an external magnetic field. Thus, although the form of the order parameters is the same as in the FF and LO states, the symmetry of the system is different since time reversal invariance is not explicitly broken. In Eq. \eqref{eq:expansion} we denoted by $\Delta_0(\bm r)$ the order parameter for an uniform SC, and $\Delta_{\pm{\bm Q}}(\bm r)$ are the components close to the ordering wave vector $\pm {\bm Q}$. A multi-component order parameter of the form of Eq.\eqref{eq:expansion} can describe the following phases: a)  a uniform SC state if  $\Delta_0\neq 0$ and $\Delta_{\pm{\bm Q}}=0$, b) an FF-type  state if $\Delta_0=\Delta_{-{\bm Q}}=0$ but $\Delta_{{\bm Q}}\neq 0$,  c) a PDW  SC state if $\Delta_0=0$ and $|\Delta_{\bm Q}|=|\Delta_{-{\bm Q}}|\neq0$, and d) a striped SC phase if $\Delta_0\neq 0$ and $|\Delta_{\pm {\bm Q}}|\neq 0$. The PDW SC is a time-reversal invariant LO state (with wave vector $\bm Q$), and the FF state is a SC (with wave vector $\bm Q$) with a spontaneously broken time-reversal invariance. The PDW state has two complex order parameters, $\Delta_{\pm {\bm Q}}(\bm r)$, whereas the FF state has  one complex order parameter field $\Delta_{\bm Q}(\bm r)$.

The recent interest in the PDW state was motivated as a symmetry-based explanation\cite{Berg-2007} of the dramatic layer decoupling effect discovered first in {\LBCO} \cite{Li-2007,Tranquada-2008} and shortly thereafter in underdoped La-based cuprate SC in magnetic 
fields.\cite{hucker-2013,Stengen-2013,Schafgans-2010} Since then there has been an extensive effort on exploring the phenomenological consequences of this state and of its possible microscopic origins. 
A Landau-Ginzburg 
theory of the PDW state gives a compelling
description of the observed phenomenology of {\LBCO},\cite{Li-2007,Berg-2007,agterberg-2008,Berg-2009} and of {\LSCO} 
in magnetic fields as a consequence of the symmetry of this SC state in a material with an LTT crystal structure.
 One important feature of the PDW state is that it is naturally intertwined with a charge ordered state with wave vector ${\bm K}=2{\bm Q}$
  (the same relation that is seen between the CDW order and the SDW order in LBCO). Several important predictions follow from the structure of 
  the Landau-Ginzburg theory.\cite{Berg-2009b,agterberg-2008,fradkin-2014} One of them   is the existence of three types of topological excitations: 
  1) SC vortices, 2) half-vortices bound to single dislocations of the charge order, and 3) double dislocations of the charge order.  
  An extension of the Kosterlitz-Thouless theory to a system with these topological excitations leads to the prediction of a complex phase 
  diagram which embodies the notion of intertwined orders resulting from the proliferation of the topological defects.\cite{Berg-2009b} 
  An interesting result is the prediction of a possible charge $4e$ SC state in the phase diagram and a $hc/(4e)$ flux quantization in the PDW state. 
  An extension of the Landau-Ginzburg theory with finite magnetic fields has shown that PDW states may arise in the vortex cores of the mixed phase,\cite{agterberg-2014} as suggested by the observation of the layer-decoupling effect\cite{Schafgans-2010} in the same regime where earlier neutron 
  scattering experiments found a magnetic-field induced SDW order.\cite{lake-2002} 

On the other hand, in spite of some significant recent progress, the microscopic origin of PDW-type states remains a challenging problem. 
At the root of these problems is the fact that SC states that condense at finite wave vector cannot arise as weak-coupling instabilities of a Fermi liquid and, hence,  cannot be reliably described by mean-field BCS theory. 
Thus, although mean-field theory does allow for PDW states to occur,\cite{Loder-2010,Loder-2011,Yang-2008b} the required critical couplings are 
typically larger than the bandwidth. Hence, these SC states can arise more naturally in regimes of strong correlation. 
Indeed,  an infinite projected entangled paired states (iPEPS) simulation of the 2D $t-J$ model finds that the uniform $d$-wave SC is 
essentially degenerate with a PDW state and with a striped SC over a wide range of dopings and of the ratio $t/J$,\cite{corboz-2014} 
(and in variational Monte-Carlo simulations.\cite{Himeda-2002,Raczkowski-2007}) 
This  result suggests that  strongly correlated systems have a strong tendency to exhibit intertwined 
orders and that the PDW state may occur more broadly than was previously anticipated.\cite{fradkin-2014}
However, a more recent density-matrix renormalization group (DMRG) simulation of a (frustrated) $t-J$ model has not found evidence for such 
ground state degeneracy.\cite{Dodaro-2016} PDW states have been proposed to be the ground state of models of strongly correlated systems, 
based on the concept of Amp{\'e}rean pairing,\cite{lee-2014} as  an explanation the pseudo-gap features found in angle-resolved photoemission 
experiments,\cite{ARPES} PDW states have been shown to arise as instabilities of spin-triplet nematic Fermi fluids.\cite{Soto-Garrido-2014} 
PDW SC states have been shown to be the ground state of 1D systems such as the Kondo-Heisenberg chain\cite{Berg-2010} and extended  
Hubbard-Heisenberg  models of 2-leg ladders for certain electronic densities,\cite{Jaefari-2012} and in a  quasi-1D model of coupled 2-leg
 ladders.\cite{Soto-Garrido-2015} Using a different approach, using the spin-fermion model Wang and collaborators \cite{Wang-2014,Wang2-2015,Wang-2015} 
 provided evidence for the coexistence of CDW and PDW orders in underdoped cuprates.

In this paper we  return to the problem of finding additional experimentally testable tell-tale predictions that may unambiguously
 identify the PDW SC state. Although the recent detection of a real-space modulation of the superfluid density in {\BSCCO} deep in the $d$-wave 
 SC state by  scanned Josephson tunneling  microscopy \cite{Hamidian-2016} is an encouraging development, there is so far no direct 
 experimental evidence of the PDW state. An observation of a half-vortex (or of the predicted anomalous flux quantization) 
 would certainly give strong indication of the observation of this state. Here we will propose a different  way to detect the PDW state.
One possibility is to study the signatures of the PDW in  Raman spectroscopy, in momentum-resolved electron energy loss spectroscopy, or in resonant inelastic X-ray scattering. For conventional superconductors, 
at low temperatures the SC amplitude mode 
(also referred as Higgs mode in similarity to the Higgs boson in high-energy physics) can be present.\cite{Varma-2015}
For conventional superconductors (such as NbSe$_2$) where the SC order parameter is uniform, the experimental signatures agree with the 
theoretical results.\cite{sooryakumar-1980,measson-2014,Littlewood-1982,Browne-1983,cea-2014}

In this paper we study the amplitude (Higgs modes) fluctuations of the PDW order parameter. Since the PDW state has two complex order parameters we expect to see two Higgs modes. In addition, in a phase in which the uniform $d$-wave state and the PDW phase coexist (a ``striped SC") a new Higgs mode should appear.  Thus, the pattern of observed Higgs modes is a signature of these SC states.  In addition,  in contrast to the usual uniform SC state, the FS in a PDW state  is not completely gapped and instead presents  Fermi pockets of Bogoliubov quasiparticles. On the other hand, since the FS surface in the PDW state remains largely gapless, expecting more decay channels for the Higgs modes one would naively expect that the damping of the Higgs modes will be stronger than in the case of the uniform SC state. However, we show below that this is not the case. As will be discussed in the following sections, the predicted  damping is expected to have a comparable threshold for the PDW SC as for a uniform SC state.

Since the PDW states cannot be studied within a weak coupling theory, here will  use instead the following approach. 
We will postulate the existence of a PDW SC state (the same with the other states) described by a quantum order parameter field 
$\Delta_{\pm {\bm Q}}({\bm r},t)$. We will assume that the quantum dynamics of this order parameter has an effective 
Lagrangian (without damping)  with dynamic critical exponent $z=1$. This quantum fluctuating field will be coupled to fermion bilinears for 
Cooper pairs with the requisite center-of-mass momentum. To simplify matters we will assume that the Fermi surface of the fermions is circular 
(although it is straightforward to generalize to other more physically-motivated cases). In particular, no assumptions on nesting will be made. 
Theories of this type are commonly used in studies of quantum criticality in metals (see, e.g. Ref.[\onlinecite{sachdev-1999}]). 
 Theories of this type have several parameters that describe the normal and the condensed phases, as well as a stiffness and a Yukawa 
coupling between the fermions and the order parameter field. These details, and others, will be specified in the subsequent sections. 

Within this framework, we study the energies and damping of the Higgs modes for a PDW state. To gain some intuition of the problem, we first analyze the case of an FF state (i.e., with only one ordering momentum ${\bm Q}$), and extend our approach to the PDW state. Our key findings are as follows. The energies of the two Higgs modes are split from $2|\Delta_{\bm Q}^{(0)}|$, i.e., twice of value of the mean-field gap. The magnitude of the splitting is given by the bi-quadratic coupling of  $\Delta_{\pm\bm Q}$, and  is generally comparable to $\Delta_{\bm Q}^{(0)}$. These two modes can decay via scattering into  bosonic collective Goldstone modes or fermionic quasiparticles. However, by gauge invariance the former process can only occur via a gradient coupling interaction, and hence is weak. The major contribution to the decay comes from scattering into fermionic quasiparticles. We find that this process has a threshold at approximately $2|\Delta_{\bm Q}^{(0)}|$, and, as a result, only one of the two Higgs modes is significantly damped, and the other remains sharp. Since in cuprate systems, a uniform (d-wave) SC component always exists at sufficiently low temperatures, we also discuss the properties of the PDW Higgs modes in the presence of a uniform SC order. In particular, we computed their energies and damping in two limiting cases when the mixing between PDW Higgs modes and uniform SC Higgs mode is weak, and we find that in both limits, it remains true that only one of the two PDW Higgs mode is significantly damped.

Collective modes of FFLO-type phases have been studied theoretically in imbalanced Fermi gases (of interest in ultra-cold atomic gases) in isotropic backgrounds.\cite{Radzihovsky-2009,Radzihovsky-2011} In such systems there is a delicate interplay between rotational and translational collective modes which was studied in detail in these references. Here, instead,  we are interested in PDW-type phases which are strongly coupled to the underlying (square) lattice and orientational collective modes are strongly gapped. Thus, unlike similar problems in cold-atomic gases, the orientation of the ordering wave vector of the PDW (and FF) states is fixed and does not fluctuate. This difference changes significantly the physics of the collective modes, particularly of the Goldstone modes. In addition,  the PDW SC is charged and couples to the electromagnetic field (through the Higgs mechanism) and, as expected, the SC Goldstone mode is absent (`Higgsed'). Also, the  damping of the collective modes was not considered in Refs.[\onlinecite{Radzihovsky-2009,Radzihovsky-2011}], while it plays a significant role in the systems that we are interested in here.

This paper is organized as follows. In Sec. \ref{sec:FF} as a warm-up example we study first the simpler Fulde-Ferrell state. We study in detail the amplitude fluctuations of the order parameter (Higgs mode), computing its energy and damping. In Sec. \ref{sec:PDW} we generalize the model of Sec. \ref{sec:FF} to two superconducting order parameters with $\bm Q$ and $-\bm Q$ momenta, i.e., the PDW state. We show the appearance of two Higgs modes, one of which is always damped. In Sec. \ref{sec:coexistence} we study the fate of PDW Higgs modes in the phase where the PDW state coexists with the usual uniform SC state. In Sec. \ref{sec:conclusions} we present our conclusions and final remarks. The details of some calculations are presented in the Appendix.

\section{Higgs mode in a Fulde-Ferrel state}
\label{sec:FF}

As a warm-up, we first study the energy and damping of the Higgs mode in a Fulde-Ferrell (FF) state, which was theoretically proposed as a superconducting state stabilized in the presence of a magnetic field.  Such a state is characterized by a single SC order parameter $\Delta_{\bm Q}$ that carries a finite momentum $\bm Q$. An FF state is very similar to the PDW state, and the only difference for the PDW state is that it has SC order parameters with both $\pm\bm Q$, and hence time-reversal is preserved. The analysis of the Higgs mode in an FF state serves as a good starting point for that in a PDW state. To our knowledge, the Higgs mode for an FF state has not be analyzed before.

\begin{figure}[t]
\centering
\includegraphics[width=1.1\columnwidth]{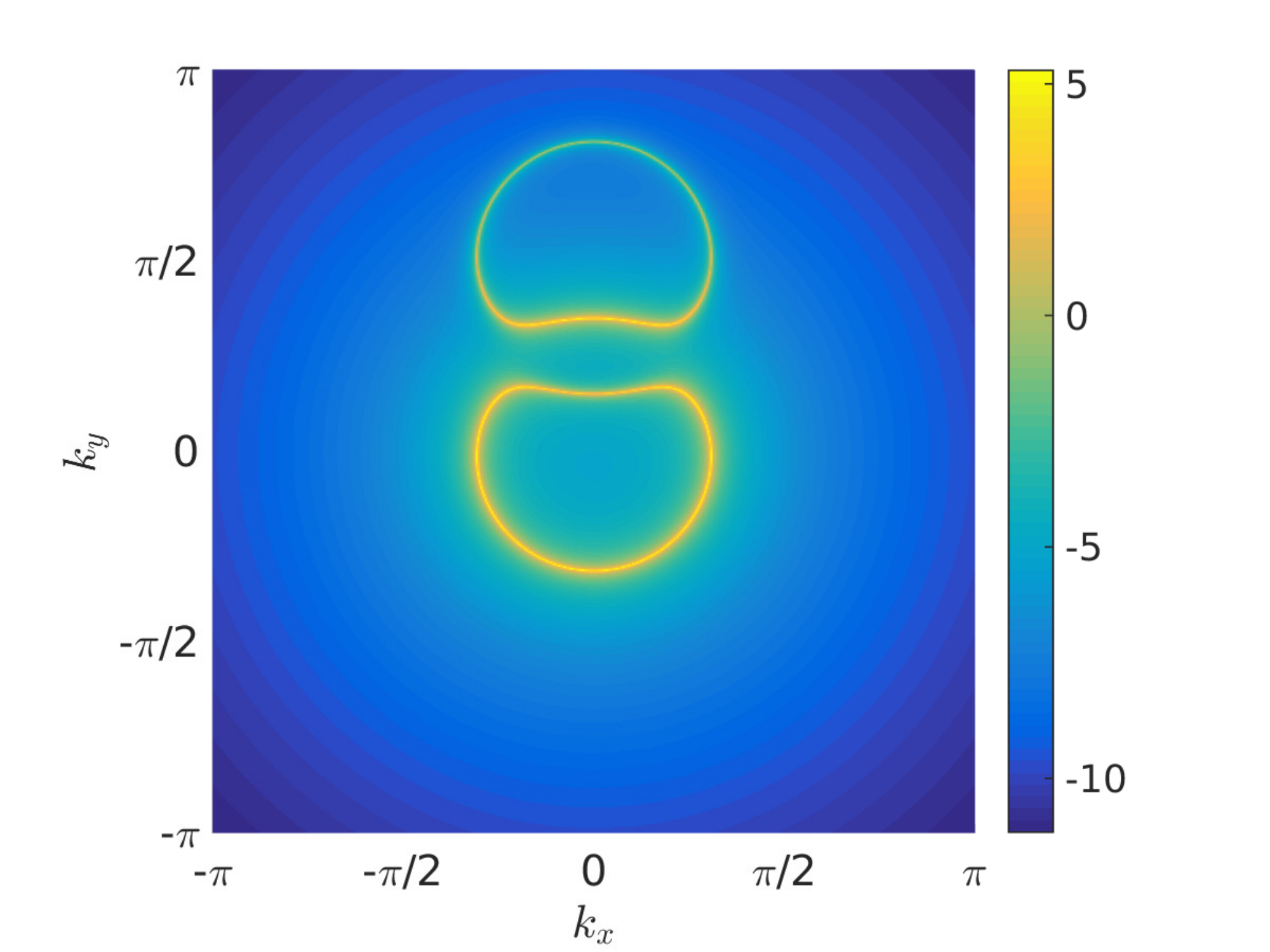}
\caption{(Color online) The spectral function for the FF state with $\bm Q=(0,\pi/2)$.}
\label{fig:spectralFF}
\end{figure}

The order parameter couples with fermions via a bilinear term
\begin{align}
\mathcal{H}_{\Delta_{\bm Q}}=\Delta_{\bm Q}^*\psi_{\bm k+\frac{\bm Q}{2}}\psi_{-\bm k+\frac{\bm Q}{2}}+\text{c.c},
\label{fermion}
\end{align}
where, for simplicity, we have kept an implicit spin structure on the fermions (which forms a singlet). Unlike the usual superconducting order parameter, an FF order parameter does not gap the full FS, but rather renormalizes the original FS into pockets. In Fig.\ \ref{fig:spectralFF}, we show the simulation of the spectral function $A(\bm k, \omega=0)={\rm Im}~G(\bm k, \omega=0)$ for a circular FS and an FF order parameter with $\bm Q=(0,\pi/2)$. On a pocket, the nature of the quasiparticle excitations continuously varies from electron-like to hole-like. 
In such a state, there are two collective modes: one Goldstone mode corresponding to the fluctuation of the phase of $\Delta_{\bm Q}$ (which gets ``eaten" by the electromagnetic field), and another Higgs mode corresponding to the fluctuation of the magnitude of $\Delta_{\bm Q}$. The two modes are decoupled, as protected by gauge invariance, and we only focus on the Higgs mode.
 
 We begin with a generic form of the Lagrangian for the FF state,
 \begin{align}
\mathcal{L}=&\kappa_0|\partial_\tau\Delta_{\bm Q }|^2+\kappa_1|\nabla\Delta_{\bm Q}|^2 +r|\Delta_{\bm Q}|^2+ u|\Delta_{\bm Q}|^4,
\label{eq:freeenergy0}
\end{align}
where $r<0$, $\tau$ is the imaginary time (for convenience with signs) and the $\nabla$ term captures the slow-varying component of the FF order parameter.
Minimizing Eq.\eqref{eq:freeenergy0}, we have $\Delta_{\bm Q}=\sqrt{-r/2u}e^{i\phi}\equiv\Delta_{\bm Q}^{(0)}e^{i\phi}$, where $\phi$ is an arbitrary phase. The Higgs mode corresponds to the longitudinal fluctuation (i.e., with $\phi$ fixed) of $\Delta_{\bm Q}$ around this value. Expanding the Lagrangian in terms of $\delta\Delta_{\bm Q}\equiv \Delta_{\bm Q}-\Delta_{\bm Q}^{(0)}$, we have
\begin{align}
\delta\mathcal{L}=&(r-\kappa_0\nu^2+\kappa_1{\bm q}^2+6u|\Delta_{\bm Q}^{(0)}|^2)|\delta\Delta_{\bm Q}(\bm{q},\nu)|^2\nonumber\\
=&(4u|\Delta_{\bm Q}^{(0)}|^2-\kappa_0\nu^2+\kappa_1{\bm q}^2)|\delta\Delta_{\bm Q}(\bm{q},\nu)|^2,
\end{align}
where $\nu$ is the frequency and $\bm {q}$ is the momentum deviation from the mean-field value ${\bm Q}$. We find that the dispersion relation of the Higgs mode is given by
\begin{align}
\nu(\bm q)=\sqrt{4u|\Delta^{(0)}_{\bm Q}|^{2}/\kappa_0+\kappa_1{\bm q}^2/\kappa_0}.
\end{align}
For optical probes such as Raman scattering, we are primarily concerned with $\bm q=0$, and in this case $\nu=2|\Delta_{\bm Q}^{(0)}|\sqrt{u/\kappa_0}$.

\begin{figure}
\includegraphics[width=0.6\columnwidth]{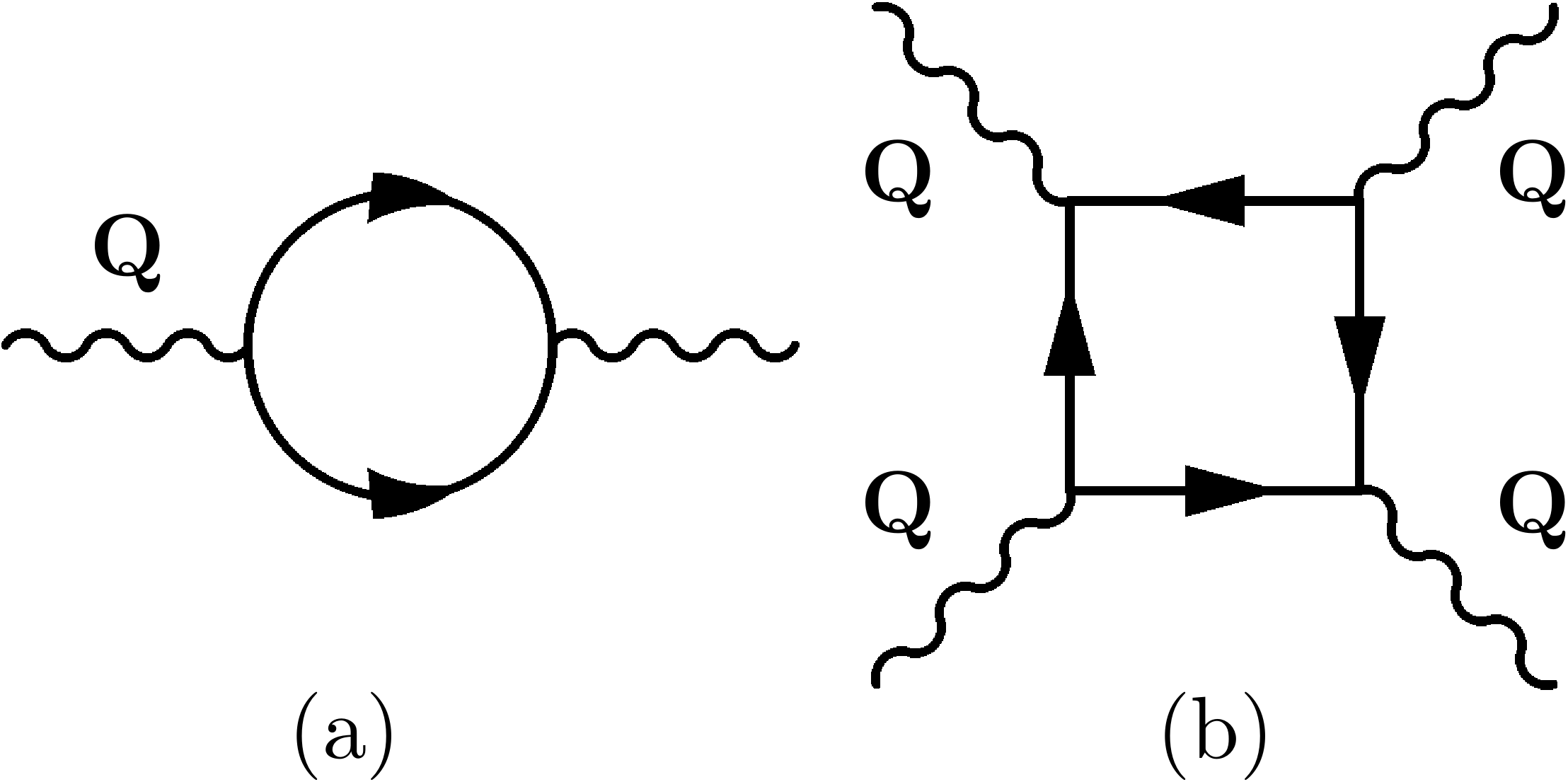}
\caption{Feynman diagrams that are relevant to coefficients $\kappa_0$ [Panel (a)] and $u$ [Panel (b)].}
\label{diag1}
\end{figure}

Furthermore, if we assume that the FF state comes from a Fermi surface instability, the coefficients $\kappa_0$ and $u$ can be computed via diagrams (see Fig.\ \ref{diag1}) by integrating out fermions. Note that this procedure is controlled only at weak coupling as it neglects fluctuations in all other channels. However, an FF state (and a PDW state we shall consider in the next section) usually does not emerge as a weak coupling instability. With this important caveat in mind, we proceed to at least get an estimate of the Higgs energy. From the diagrams in Fig.\ \ref{diag1}, we obtain
\begin{align}
\kappa_0=&\frac{1}{4}\frac{\partial^2}{\partial\Omega_m^2}\[\int_{\bm k,\omega_m}\frac{1}{i(\omega_m+\Omega_m)-\xi_{\bm {k}+\bm Q}}\frac{1}{-i\omega_m-\xi_{-\bm k}}\]_{\Omega_m=0} \nonumber\\
u=&\frac{1}{4}\int_{\bm k,\omega_m}\frac{1}{[i\omega_m-\xi_{\bm {k}+\bm Q}]^2}\frac{1}{[-i\omega_m-\xi_{-\bm k}]^2}.
\label{eq:kappaandu}
\end{align}
where $\omega_m$ is the Matsubara frequency corresponding to imaginary time, and we have used the shorthand notation $\int_{\bm k,\omega_m}=\int{d^dk d\omega}/{(2\pi)^{d+1}}$. Note that, in the spirit of the Ginzburg-Landau expansion,  we use the normal state Green's functions and order parameters appear as vertices.
After taking the derivative and integrating by parts in the expression for $\kappa_0$, we find that $\kappa_0\equiv u$ for \emph{any} band structure $\xi_{\bm k}$. The value of $u$ can be estimated by noticing that Eq.\eqref{eq:kappaandu} does not contain any IR divergence because of the double poles. As a result, by dimensional analysis $\kappa_0=u\sim1/(v_FE_F)$ in 2D.

 Thus, we find in this case that
\begin{equation}
 \nu=2|\Delta_{\bm Q}^{(0)}|.
 \label{eq:higgsFFenergy}
\end{equation}
The same result can be obtained from solving the self-consistent equations (which technically speaking is also only controlled in weak coupling) for the Higgs and Goldstone modes in the FF state, which we detail in Appendix \ref{app:FF}.

\begin{figure}
\includegraphics[width=0.4\columnwidth]{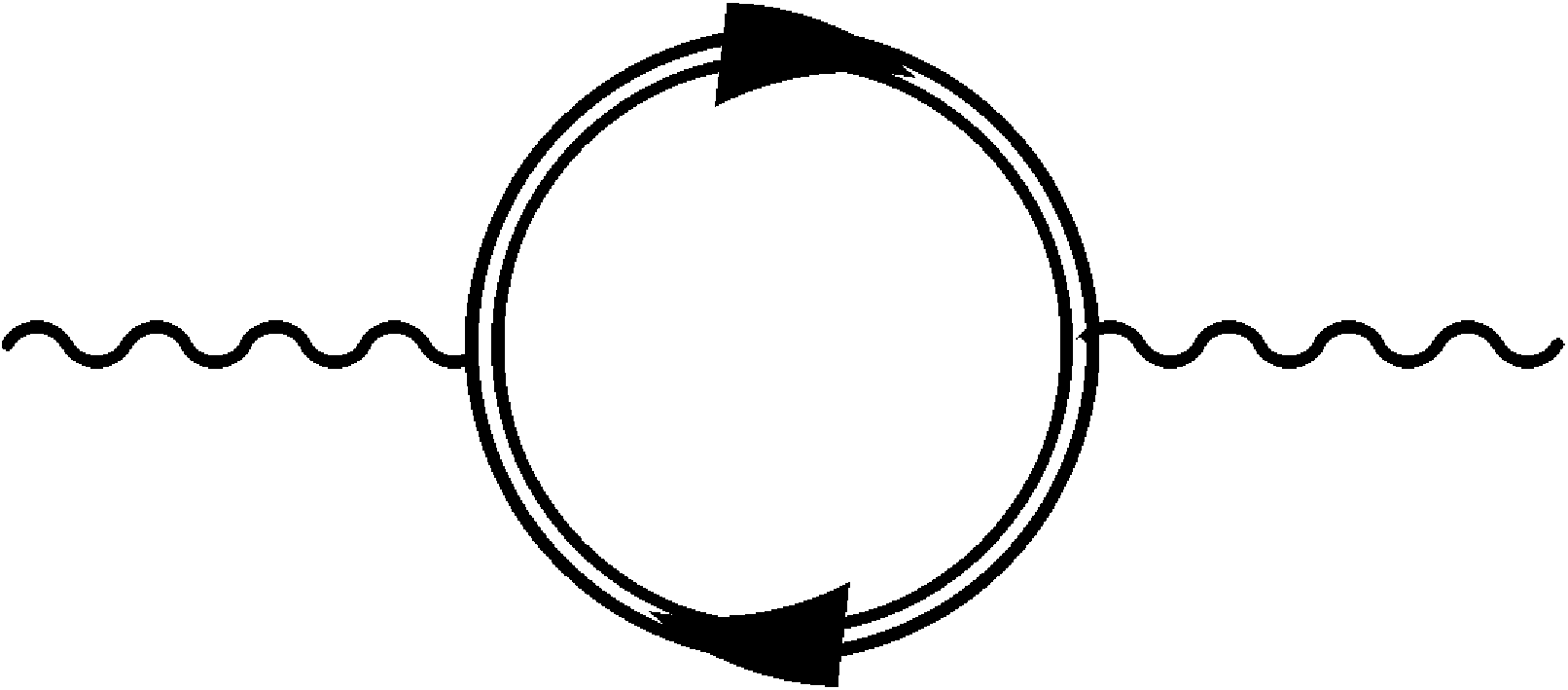}
\caption{Feynman diagrams for the damping of the Higgs mode.}
\label{dresdiag1}
\end{figure}

Next we discuss the damping (decay) of the Higgs mode. For a neutral superfluid, the Higgs mode can weakly [via a gradient coupling, required by gauge invariance (see, e.g., Refs.\ \onlinecite{Watanabe,Arovas-2011})] decay into two massless Goldstone modes, and the scattering process is associated with the masses of the Higgs mode and the Goldstone mode by a Ward identity. 
However, in a superconductor, as we said, the Goldstone 
mode is absent, since it gets absorbed (``eaten'') by the electromagnetic field, making the latter gapped at the plasmon frequency via the Anderson-Higgs mechanism. In this case, we will only consider the damping of the Higgs mode via decaying into two fermionic quasiparticles.
Such a process can be evaluated by computing the imaginary part of the particle-particle bubble, shown in Fig.\ \ref{dresdiag1}. Again, this calculation is strictly speaking only well-controlled at weak coupling. In this diagram, the wavy lines represent the Higgs mode, i.e., the amplitude fluctuation around the mean-field value of $\Delta_{\bm Q}$, while the fermion double lines are the ones renormalized by the mean field of $\Delta_{\bm Q}$. This is equivalent to summing up the full series of diagrams like those in Fig.\ \ref{diag1} while in each diagram keeping two external legs to be the Higgs mode and the rest as the mean field.
The fermionic Green's function in the presence of the mean field $\Delta_{\bm Q}$ is given by
\begin{widetext}
\begin{align}
G({\bm k},\omega_m)=\frac{\Delta_{\bm Q}^{(0)}\tau_1+(\xi_{-\bm{k}+\bm Q}+\xi_{\bm k})\tau_3/2+[i\omega_m+(\xi_{-\bm{k}+\bm Q}-\xi_{\bm k})/2]}{(\omega_m-i\xi_{-\bm{k}+\bm Q})(\omega_m+i\xi_{\bm k})+|\Delta_{\bm Q}^{(0)}|^2},
\label{eq:GFDelta}
\end{align}
\end{widetext}
where $\tau$'s are Pauli matrices in the Nambu space, and as before we have set both the mean-field order parameter $\Delta_{\bm Q}^{(0)}$ and the Higgs mode $\delta\Delta_{\bm Q}$ as real (indicated by $\tau_1$). 
The polarization operator is thus given by
\begin{align}
K(\Omega_m)=&-\frac{1}{2}\int_{\bm k,\omega_m}\tr\[G({\bm k},\omega_m+\Omega_m)\tau_1G({\bm k},\omega_m)\tau_1\]\nonumber\\
=&\int_{\bm k}\frac{1}{\sqrt{4|\Delta_{\bm Q}^{(0)}|^2+(\xi_{\bm{k}}+\xi_{-\bm{k}+\bm Q})^2}}\nonumber\\
&\times\frac{ (\xi_{\bm{k}}+\xi_{-\bm{k}+\bm Q})^2}{\Omega_m^2+4|\Delta_{\bm Q}^{(0)}|^2+(\xi_{\bm{k}}+\xi_{-\bm{k}+\bm Q})^2}
\label{eq:bubbleFF}
\end{align}
The $\bm k$ integral is done in the range satisfying $\xi_{\bm{k}}\xi_{-\bm{k}+\bm Q}+|\Delta_{\bm Q}^{(0)}|^2>0$. This is a result of the frequency integral we performed in the second line, which requires the poles in $\Omega_m$ to be located in opposite half complex planes.

To compute the damping of the modes, we need to evaluate the imaginary part of Eq. \eqref{eq:bubbleFF}. Doing the analytical continuation to real frequencies $i\Omega_m\to\nu+i\eta$, the imaginary part of Eq.\eqref{eq:bubbleFF} reads:
\begin{align}
&\pi\int_{\bm k}\frac{(\xi_{\bm{k}}+\xi_{-\bm{k}+\bm Q})^2\delta\left(\nu^2-4|\Delta_{\bm Q}^{(0)}|^2-(\xi_{\bm{k}}+\xi_{-\bm{k}+\bm Q})^2 \right)}{\sqrt{4|\Delta_{\bm Q}^{(0)}|^2+(\xi_{\bm{k}}+\xi_{-\bm{k}+\bm Q})^2}} 
\label{eq:2.9}
 \end{align}
Defining $\xi_{\pm}=\xi_{\bm{k}}\pm \xi_{-\bm{k}+\bm Q}$ we can write the previous integral as:
 \begin{align}
&\int\frac{d\xi_+d\xi_-}{2}\frac{\xi_+^2}{\sqrt{4|\Delta_{\bm Q}^{(0)}|^2+\xi_+^2}} \delta\left(\nu^2-4|\Delta_{\bm Q}^{(0)}|^2-\xi_+^2 \right)
 \end{align}
where integral over $\xi_-$ ranges from $-|\nu|$ to $|\nu|$ due to the restriction $\xi_{\bm{k}}\xi_{-\bm{k}+\bm Q}+|\Delta_{\bm Q}^{(0)}|^2>0$. 
Finally we have:
 \begin{align}
\Im(K)&=\int\frac{d\xi_+d\xi_-}{2}\frac{\xi_+^2}{\sqrt{4|\Delta_{\bm Q}^{(0)}|^2+\xi_+^2}} \delta\left(\nu^2-4|\Delta_{\bm Q}^{(0)}|^2-\xi_+^2 \right)
\nonumber\\
&=\frac{1}{2}\frac{\sqrt{\nu^2-4|\Delta_{\bm Q}^{(0)}|^2}}{|\nu|}\int_{-|\nu|}^{|\nu|}d\xi_-=\sqrt{\nu^2-4|\Delta_{\bm Q}^{(0)}|^2},
\label{eq:dampingFF}
 \end{align}
 given that $\nu\geq 2|\Delta_{\bm Q}^{(0)}|$, and for $\nu\leq 2|\Delta_{\bm Q}^{(0)}|$, $\Im(K)=0$.


We can see from Eq.\eqref{eq:higgsFFenergy} and Eq.\eqref{eq:dampingFF} that the Higgs mode in the FF state is right at the threshold energy, and therefore is not damped. This may sound trivial, since it is the same as the damping of the Higgs mode for a uniform SC state. However, the important difference is that in a uniform SC state the FS is completely gapped, while in the FF state, there exist abundant gapless fermions which form pockets.
Surprisingly, we have found that the existence of the gapless fermionic quasiparticles cannot act as a decay channel for the Higgs mode. The main reason is kinematics. As seen from the above calculations, the damping of the Higgs mode is tied to fermions with momenta $\bm {k}$ and ${\bm{-k}+\bm{Q}}$. The total energy of two quasiparticles made out of these two fermions are always gapped, and, as the $\delta$-function in Eq.\eqref{eq:dampingFF} suggests, is given by $E(\bm k)=\sqrt{4|\Delta_{\bm Q}^{(0)}|^2+(\xi_{\bm k}+\xi_{\bm {-k+Q}})^2}$. The two-particle energy $E$ can also be obtained by directly diagonalizing a 2-by-2 Hamiltonian 
$H(\bm k)=\Delta_{\bm Q}^{(0)}\tau_1+(\xi_{-\bm{k}+\bm Q}+\xi_{\bm k})\tau_3/2+(\xi_{-\bm{k}+\bm Q}-\xi_{\bm k})/2$ and summing the absolute values of its two eigenvalues. Since $E$ has a minimum value of $2|\Delta_{\bm Q}^{(0)}|$, the threshold frequency for damping is also $2|\Delta_{\bm Q}^{(0)}|$, as we found in Eq.\eqref{eq:dampingFF}.

\section{Higgs modes in a Pair-density-wave state}
\label{sec:PDW}
\begin{figure}[h]
\includegraphics[width=1.1\columnwidth]{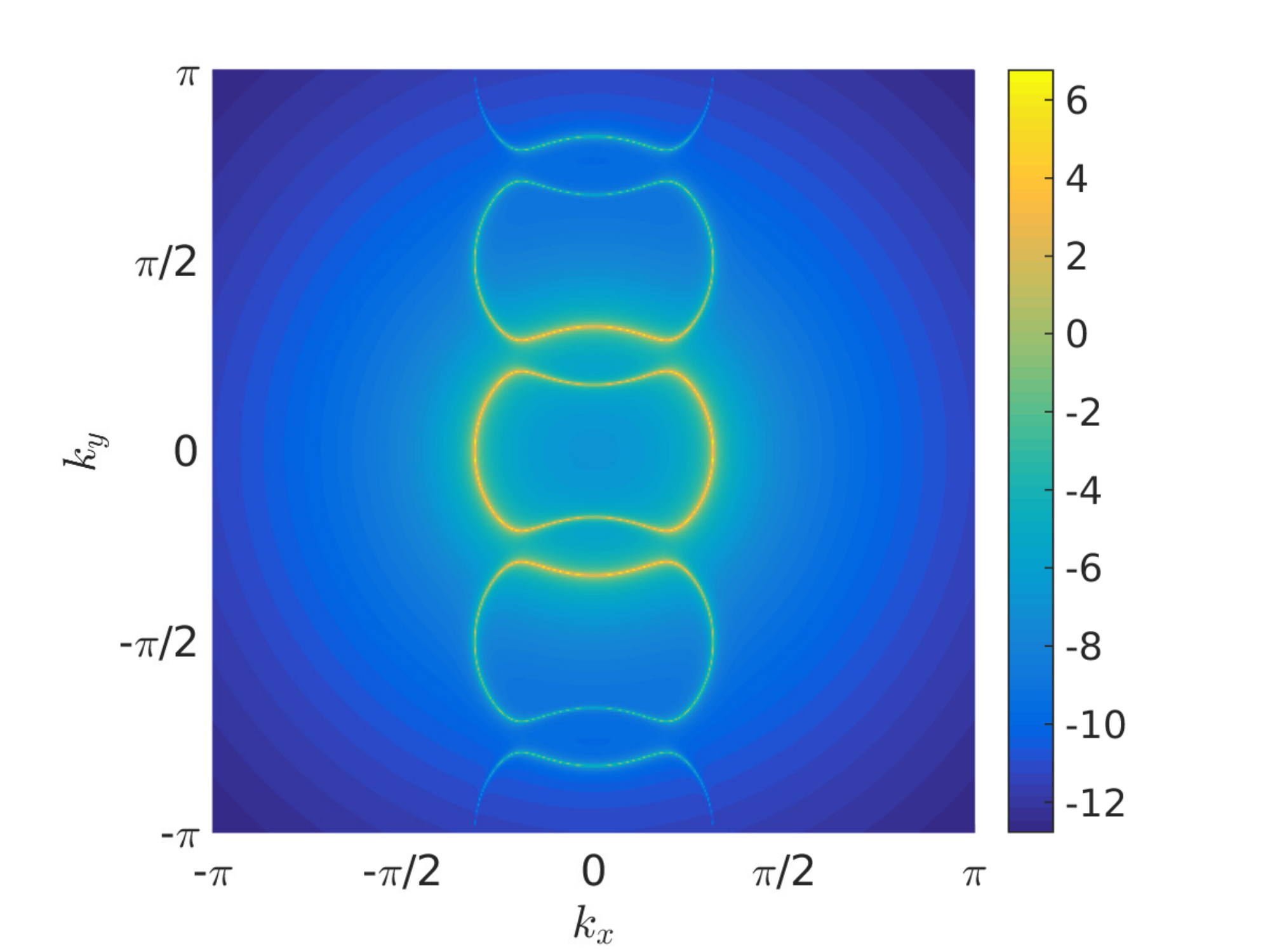}
\caption{The spectral function for a pair-density-wave state with wave vectors $\pm \bm{Q}=\pm (0,\pi/2)$. The pockets are clearly visible.}
\label{fig:pockets}
\end{figure}
We now analyze the Higgs modes in a PDW state. As we said, the main difference with the FF state is that, for a PDW state,  SC order parameters with both wave vectors $\pm \bm Q$ are present. Like in the FF state, in the PDW state the fermionic quasiparticles remain gapless and form pockets. We show in Fig.\ \ref{fig:pockets} such pockets with the simulation of the spectral function in a PDW state.

As we did for the FF state, we consider a generic form of the Lagrangian for the PDW state. In this case we have two complex order parameters $( \Delta_{\bm Q},\Delta_{-{\bm Q}})$ and up to quartic order the Lagrangian is given by: 
\begin{align}
\mathcal{L}=&\kappa_0\left(|\partial_\tau\Delta_{\bm Q }|^2+|\partial_\tau\Delta_{-\bm Q}|^2\right) 
+\kappa_1\left(|\nabla\Delta_{\bm Q}|^2+|\nabla\Delta_{-\bm Q}|^2\right) \nonumber\\
&+ r\big( |\Delta_{\bm Q}|^2+|\Delta_{-\bm Q}|^2\big) \nonumber\\
&+ u\left(|\Delta_{\bm Q}|^4+|\Delta_{-\bm Q}|^4\right)+\gamma|\Delta_{\bm Q}|^2|\Delta_{-\bm Q}|^2
\label{eq:freeenergy}
\end{align}
For $r<0$, the system enters an ordered state. Whether both $\Delta_{\pm {\bm Q}}$ have nonzero expectation values depends on the interplay between $u$ and $\gamma$. To see this, we neglect the spatial and temporal dependence in Eq.\eqref{eq:freeenergy}:
\begin{align}
\mathcal{U}=&r\left(|\Delta_{\bm Q}|^2+|\Delta_{-\bm Q}|^2\right)+u\left(|\Delta_{\bm Q}|^2+|\Delta_{-\bm Q}|^2\right)^2\nonumber\\
&+(\gamma-2u)|\Delta_{\bm Q}|^2|\Delta_{-\bm Q}|^2.
\label{free2}
\end{align}
When minimizing this free energy, the first two terms fix the mean-field value of $|\Delta_{\bm Q}|^2+|\Delta_{-\bm Q}|^2$. For $\gamma\geq2u$, the last term fixes one of $\Delta_{\pm {\bm Q}}$ to be zero, and the resulting ground state is an FF state. For $\gamma<2u$, the last term favors that $|\Delta_{{\bm Q}}|=|\Delta_{-{\bm Q}}|\neq 0$, and the ground state is a PDW state. Furthermore, when $\gamma<-2u$, the free energy Eq.\eqref{free2} becomes unbounded as the quartic terms can be made arbitrarily negative for a large $\Delta_{{\bm Q}}=\Delta_{-{\bm Q}}$. In this case, to find the ground state, sixth order terms need to be included, and the transition into a PDW state becomes first order. For the following we only focus on the case where $-2u<\gamma<2u$. In this case the saddle point solution for the Lagrangian defined above is $\Delta_{\pm\bm Q}=\sqrt{-r/(2u+\gamma)}e^{i\phi_{\pm\bm Q}}\equiv\Delta_{\pm\bm Q}^{(0)}e^{i\phi_{\pm\bm Q}}$, where $\phi_{\pm\bm Q}$ is an arbitrary phase and we have defined:
\begin{equation}
 \Delta^{(0)}_{\bm Q}=\Delta^{(0)}_{-\bm Q}=\sqrt{\frac{-r}{(2u+\gamma)}}
 \label{eq:SPsol}
\end{equation}
The Higgs modes correspond to the longitudinal fluctuations (i.e., with $\phi_{\pm\bm Q}$ fixed) of $\Delta_{\pm\bm Q}$ around the saddle point value in Eq. \eqref{eq:SPsol}. Expanding the Lagrangian in terms of $\delta\Delta_{\pm\bm Q}\equiv \Delta_{\pm\bm Q}-\Delta_{\pm\bm Q}^{(0)}$ we have
\begin{align}
\delta\mathcal{L}=&\left(r-\kappa_0\nu^2+\kappa_1|\bm{q}|^2\right)\left(\delta\Delta_{\bm Q}^2+\delta\Delta_{-\bm Q}^2\right)\nonumber\\ 
&+6u\left((\Delta^{(0)}_{\bm Q})^2\delta\Delta_{\bm Q}^2+
(\Delta^{(0)}_{-\bm Q})^2\delta\Delta_{-\bm Q}^2\right)\nonumber\\
&+\gamma\left((\Delta^{(0)}_{\bm Q})^2\delta\Delta_{-\bm Q}^2+(\Delta^{(0)}_{-\bm Q})^2\delta\Delta_{\bm Q}^2\right)\nonumber\\
&+2\gamma\left(\Delta^{(0)}_{\bm Q}\Delta_{-\bm Q}^{(0)}\delta\Delta_{-\bm Q}\delta\Delta_{\bm Q}+\text{h.c}\right),
\label{eq:fluctuations}
\end{align}
where we have switched to momentum/frequency space. Substituting the results in Eq. \eqref{eq:SPsol} we obtain,
\begin{align}
\delta\mathcal{L}=&\bmb \delta\Delta_{\bm Q} & \delta\Delta_{-\bm Q} \emb M \bmb \delta\Delta_{\bm Q} \\ \delta\Delta_{-\bm Q} \emb 
\label{eq:fluctuations2}
\end{align}
where the matrix $M$ is given by:
\begin{align}
M=&\left(-\kappa_0\nu^2+\kappa_1|\bm{q}|^2+4u|\Delta^{(0)}|^2\right)\sigma_0+2\gamma|\Delta^{(0)}|^2\sigma_1
\label{eq:PDWMatrix}
\end{align}
and we have used $\Delta^{(0)}_{\bm Q}=\Delta^{(0)}_{-\bm Q}=\sqrt{\frac{-r}{(2u+\gamma)}}$ and $\sigma_0$ and $\sigma_1$ are the identity and the first Pauli matrix. As a generalization of that in the FF state, the dispersion of the Higgs modes in the PDW state is given by the vanishing of the determinant of  $M$, and
\begin{equation}
 \det(M)=0\implies \nu_{1,2}(\bm q)=\sqrt{\frac{(4u\pm 2\gamma) }{\kappa_0}|\Delta_{\bm Q}^{(0)}|^2+\kappa_1\bm q^2}
\end{equation}
As in the FF state, we will focus on fluctuations with $\bm q=0$. Using the condition $-2u<\gamma<2u$, the energies $\nu_{1,2}(\bm q=0)$ are both real (since we have not yet considered the damping via decaying into fermion pairs).

\begin{figure}
\includegraphics[width=0.35\columnwidth]{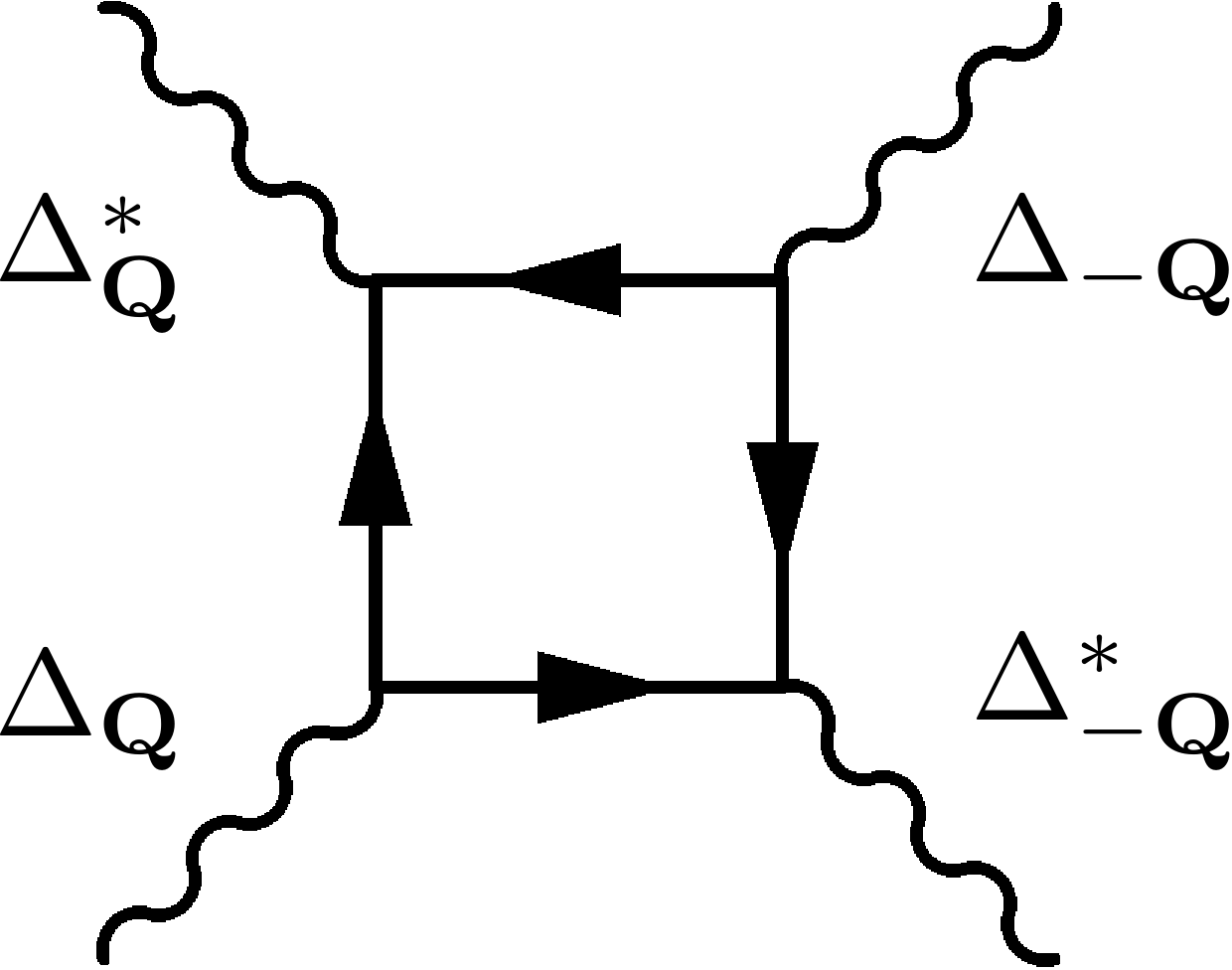}
\caption{Feynman diagram for coefficient $\gamma$.}
\label{fig2}
\end{figure}

In the case of weak coupling, the coefficients $\kappa_0$, $u$ and $\gamma$ can be computed via diagrams by integrating out fermions. The coefficients
$\kappa_0$ and $u$ are the same as those in Eq. \eqref{eq:kappaandu}, so $\kappa_0=u$.
The Higgs modes energies are thus given  by
\begin{equation}
 \nu_1=2|\Delta_{\bm Q}^{(0)}|\sqrt{1+\frac{\gamma}{2u}}\quad\text{and}\quad
 \nu_2=2|\Delta_{\bm Q}^{(0)}|\sqrt{1-\frac{\gamma}{2u}}.
 \label{eq:modesPDW}
\end{equation} 
where the coefficient $\gamma$ is given by the diagram in Fig. \ref{fig2} which corresponds to the expression:
\begin{align}
\gamma=&\frac{1}{4}\int_{\bm k,\omega_m}\frac{1}{[i\omega_m-\xi_{\bm {k}+\bm Q}]}\frac{1}{[i\omega_m-\xi_{\bm {k}-\bm Q}]}\frac{1}{[-i\omega_m-\xi_{-\bm k}]^2}.
\label{eq:gamma}
\end{align} 
The evaluation of this integral requires a detailed knowledge of the band structure. Since the integral does not contain any IR divergence, from dimensional analysis we expect $\gamma\sim u \sim 1/(v_FE_F)$ in 2D.

As usual, the energies of the Higgs modes acquire imaginary parts via decaying into bosonic collective modes or fermionic quasiparticles. Unlike the FF state, in which the Goldstone mode is absent, in a PDW state a gapless Goldstone mode does exist due to its coupling to the electromagnetic field. This is because  the PDW state breaks a $U(1)\times U(1)$ symmetry, corresponding to the phases of $\Delta_{\pm\bm{Q} }$, and has two Goldstone modes (before coupling to the electromagnetic field). The two Goldstone modes are the common phase (which carries charge $2e$) and the relative phase (which is charge neutral) of $\Delta_{\pm\bm{Q} }$. 
 Once coupled to the electromagnetic field,  the Goldstone mode corresponding to the common phase gets absorbed (``Higgsed'') and gapped at the plasma frequency. The other neutral mode, corresponding to the relative phase, is coupled to the phase of the induced CDW order parameter with ${\bm K}=2{\bm Q}$. This Goldstone mode remains gapless as long as $\bm K$ is incommensurate and the CDW does not get locked to the lattice. In this case, the PDW Higgs mode can decay into two such gapless Goldstone modes with opposite momenta, but only weakly so, as we discussed for the FF state.\cite{Watanabe, Arovas-2011} As a result, the Higgs modes have long, but finite, lifetimes, and their spectral peaks get broadened but remain well-defined. The magnitude of this small width can in principle be computed within our PDW Lagrangian, but receives an $O(1)$ correction if the dynamics of the (induced) CDW is included. For this reason we will not pursue this further, but we emphasize that this broadening from this process is small in the sense that this decaying process is via a gradient term. On the other hand, a more significant contribution to the damping of Higgs modes comes from decaying into a pair of fermionic quasiparticles. 
 
To this end, we couple the order parameters $\Delta_{\pm\bm Q}$ with the fermions. The bilinear Hamiltonian for such a coupling is given by:
\begin{align}
\mathcal{H}_{\Delta_{\bm Q}}=\Delta_{\bm Q}^*\psi_{\bm k+\frac{\bm Q}{2}}\psi_{-\bm k+\frac{\bm Q}{2}}+\Delta_{-\bm Q}^*\psi_{\bm k-\frac{\bm Q}{2}}\psi_{-\bm k-\frac{\bm Q}{2}}+\text{c.c},
\label{fermionPDW}
\end{align}
The damping of the Higgs modes is then analyzed by evaluating the imaginary part of the bubble diagram, shown in Fig.\ \ref{dresdiag1}, similar to that for the FF state. The only difference is that the fermion double line is renormalized by both $\Delta_{\pm \bm Q}$. Its Hamiltonian $H(\bm k)$ and Green's function $G(\omega_m, \bm k)$ has a much more complex matrix structure since fermions with momenta $\bm k$, $-\bm {k}+{\bm Q}$, $-\bm {k}-{\bm Q}$, $\bm {k}+2{\bm Q}$, etc., are all coupled, and the size of the matrix depends on when (and whether) this series closes under $\bm k=\bm k+2\pi$. The evaluation of such a diagram is conceptually straightforward but technically tedious. However, with what we learned for the FF state, insights into the damping can be gained without carrying out the integral. Just like Eq.\eqref{eq:2.9}, the damping of the PDW Higgs mode (say with momentum $\bm Q$) comes from an on-shell process of the Higgs mode scattering into quasiparticles made out of an electron and a hole with total momentum $\bm Q$. As before, the threshold frequency for the damping is the minimum of the total energy of such quasiparticles. For the FF state, the minimum is reached at $\xi_{\bm k}=\xi_{-\bm {k+Q}}=0$, and is $2|\Delta_{\bm Q}^{(0)}|$. For the PDW state the threshold is found by diagonalizing the full matrix Hamiltonian $H(\bm k)$ (instead of the 2-by-2 one for the FF state). However, the minimum of the two-particle energy is still found close to when $\xi_{\bm k}=\xi_{-\bm {k+Q}}=0$, and typical energies of all other fermions involved are of $O(E_F)$. For $\Delta_{\bm Q}^{(0)}/E_F\ll 1$, we can approximately reduce $H(\bm k)$ to the 2-by-2 block, and $E\approx 2|\Delta_{\bm Q}^{(0)}|$. This result is corrected by high energy fermions via $\Delta_{-\bm {Q}}^{(0)}$, and perturbation theory, the corrections are of $O((\Delta_{\bm Q}^{(0)}/E_F)^2)$. Thus, without doing any calculation, we find the damping condition for PDW Higgs mode to be
\begin{align}
\nu\geq 2|\Delta_{\bm Q}^{(0)}|\ [1+ O(\Delta_{\bm Q}^{(0)}/E_F)].
\end{align} 
Compared with the FF state, we see that the damping threshold gets shifted from  $2|\Delta_{\bm Q}^{(0)}|$. In the regime $\Delta_{\bm Q}\ll T$ where the Ginzburg-Landau Lagrangian is valid, the splitting of the Higgs modes energies is larger than the shift of the damping threshold. Therefore,  only \emph{one} of the Higgs modes gets significantly damped. This is the key result of the present work.

 Upon further lowering $T$, the quartic form of the Lagrangian, Eq.\eqref{eq:freeenergy}, becomes invalid, and the relation between the Higgs energies and the damping threshold has to be obtained from evaluating bubble diagrams with fully dressed Green's functions at finite temperatures, which is beyond the scope of our work.

\section{PDW Higgs modes in coexistence with a uniform SC order}
\label{sec:coexistence}

Experimentally, the putative PDW phase coexists with a uniform $d$-wave superconducting (SC) phase at low temperature in LBCO. It is thus interesting to consider the fate of the PDW Higgs modes in the presence of a uniform SC order. In such a coexistence phase, the PDW Higgs modes generally become mixed with the SC Higgs modes, which makes the problem in general complicated. It is helpful to consider the limiting cases, where
the uniform SC gap is either much smaller or much larger than the PDW gap. In such situations, the mixing of PDW Higgs modes with the SC one is negligible and we can focus on the PDW sector. It is beyond the scope of this work to treat the SC and PDW Higgs modes together and consider their mixings, which is again conceptually straightforward but technically tedious. 

\begin{figure}
\includegraphics[width=0.35\columnwidth]{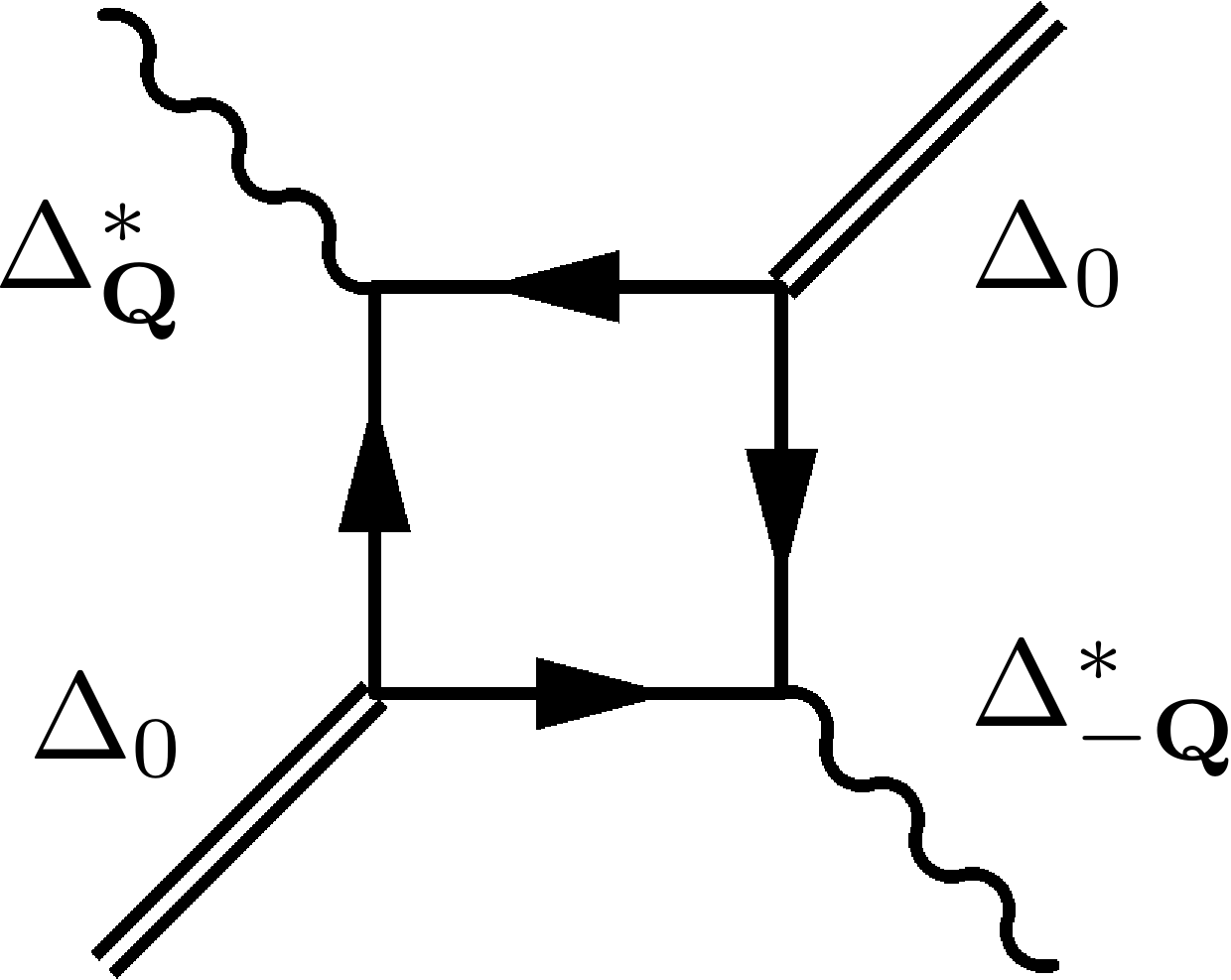}
\caption{Feynman diagram corresponding to the coefficient $\tilde\gamma$.}
\label{diag4}
\end{figure}

The generic Lagrangian for the coexistence of the  uniform SC and the PDW  has been given in Ref.\ \onlinecite{fradkin-2014}. For our purposes, we focus on the PDW sector and simply replace the SC order parameter with its mean-field value $\Delta_0$. For the SC sector, there exists the usual Higgs mode~\cite{Littlewood-1982} with frequency $\nu_{\rm SC}=2|\Delta_{0}|$. The Lagrangian for the PDW sector is given by,
\begin{align}
\mathcal{L}=&\kappa_0\left(|\partial_\tau\Delta_{\bm Q }|^2+|\partial_\tau\Delta_{-\bm Q}|^2\right)
\nonumber\\
&+r\left(|\Delta_{\bm Q}|^2+|\Delta_{-\bm Q}|^2\right)\nonumber\\
&+u\left(|\Delta_{\bm Q}|^4+|\Delta_{-\bm Q}|^4\right)+\gamma|\Delta_{\bm Q}|^2|\Delta_{-\bm Q}|^2
\nonumber\\
&+\gamma_0|\Delta_{0}|^2\left(|\Delta_{\bm Q}|^2+|\Delta_{-\bm Q}|^2\right)\nonumber\\
&+\tilde{\gamma}\left[(\Delta_{0}^*)^2\Delta_{\bm Q}\Delta_{-\bm Q}+\text{c.c.}\right].
\label{eq:freeenergy3}
\end{align}
where we have taken $\bm q=0$ as we did in the FF and PDW states.  In the coexistence phase, where $\Delta_{0}\neq0$ and $\Delta_{\pm\bm Q}\neq0$ we can expand around the saddle point solution. We see from the last term of Eq.\ \eqref{eq:freeenergy3} that the presence of a uniform SC $\Delta_{0}$ further mixes and splits the two Higgs modes. Indeed, following a similar procedure to minimize the free energy and solve for the Higgs frequencies we find,
\begin{align}
 \nu_1&=2|\Delta_{\bm Q}^{(0)}|\sqrt{1+\frac{\gamma}{2u}}\nonumber\\
 \nu_2&=2|\Delta_{\bm Q}^{(0)}|\sqrt{1-\frac{\gamma}{2u}+\frac{\tilde{\gamma}|\Delta_0|^2}{2u|\Delta_{\bm Q}^{(0)}|^2}}.
 \label{eq:modescoexistence}
\end{align} 
The coefficient $\tilde\gamma$ is given by the diagram in Fig.\ \ref{diag4}, and is expressed as
\begin{align}
\tilde\gamma=&\frac{1}{2}\int_{\bm k,\omega_m}\frac{1}{\omega_m^2+\xi_{\bm {k}}^2}\frac{1}{\omega_m^2+\xi_{\bm {-k+Q}}^2}.
\label{eq:gammatilde}
\end{align} 
Unlike Eqs.\ (\ref{eq:kappaandu}) and (\ref{eq:gamma}), the above integral is divergent in the IR which is cut by temperature, and thus $\tilde\gamma\sim1/(v_F T)$ in 2D. For $T\ll E_F$, we have $\tilde\gamma\gg \gamma,u$.

Comparing with Eq.\eqref{eq:modesPDW} and Eq.\eqref{eq:modescoexistence} we see that the existence of a uniform SC order parameter shifts the lower Higgs frequency up while keeping the other one unchanged.
However, as we said, the above approach treating the PDW Higgs modes separately from the SC Higgs mode is only valid when their energies are not close, i.e., either $\Delta_0\ll \Delta_{\bm Q}^{(0)}$ or $\Delta_0\gg \Delta_{\bm Q}^{(0)}$. 

As for the damping of the Higgs modes, when $\Delta_0\ll \Delta_{\bm Q}^{(0)}$, the quasiparticle spectrum is dominated by the PDW order parameter, and the damping threshold remains approximately $ 2|\Delta_{\bm Q}^{(0)}|$. As long as the shift in Higgs energy caused by $\Delta_0$ is small, only one Higgs mode of the two is damped, just like in the pure PDW state. 
 In the opposite limit $\Delta_0\gg \Delta_{\bm Q}^{(0)}$, the {relevant regions that may cause the damping of the PDW Higgs modes are}
gapped by uniform SC and the damping threshold becomes $2\Delta_0$.
The Higgs energies in this limit are $\nu_1\sim \Delta_{\bm Q}^{(0)}$ and  $\nu_2\sim \sqrt{\tilde\gamma/u} \Delta_0 \sim  \Delta_0\sqrt{E_F/T}\gg \Delta_0$. In this limit only $\nu_2$ is damped.
 Therefore, in both limits of the coexistence phase, it still holds true that one of the two PDW Higgs modes is damped. {We close this section by noting that even though the $d$-wave SC order parameter does not fully gap out the Fermi surface but rather leaves Dirac nodal points with gapless Bogoliubov quasiparticles, they, however,  generically cannot damp the PDW Higgs modes due to kinematics, just as in the case of the gapless quasiparticles of a pure PDW state.}


\section{Concluding Remarks}
\label{sec:conclusions}

In this work, we analyzed the properties of the Higgs modes of a pair-density-wave state, a spatially modulated superconducting state. We discussed the energy and damping of the Higgs modes, both from a phenomenological Lagrangian for the order parameters and from a self-interacting fermionic theory. We found that, even though the FS remains largely gapless (and reorganizes as pockets), the damping of Higgs mode requires exceeding a threshold frequency, which is twice the gap ($2|\Delta_{\bm Q}^{(0)}|$) due to kinematics. For the PDW state, we found that there exist two Higgs modes, and due to their mixing, the energies are split from $2|\Delta_{\pm \bm Q}^{(0)}|$. As a result, only one of the two Higgs modes is damped by the coupling to the fermionic excitations.
Motivated by the putative PDW order in the cuprate superconductors, we also discussed the fate of the PDW Higgs modes in the presence of a uniform SC order. We focused on two limiting cases where the mixing between the PDW Higgs and the uniform SC Higgs can be neglected, and analyzed the properties of the PDW Higgs modes. We found that in the coexistence phase with uniform SC, it remains that only one of the two Higgs modes is significantly damped, at least in the limit where the mixing between the PDW Higgs modes and SC Higgs mode can be neglected.

The PDW order parameters, being in the particle-particle channel,  do not directly couple to the fermion density. Just like the Higgs mode for the uniform superconductor NbSe$_2$, the PDW Higgs modes can be detected via coupling to a CDW collective mode. Since the PDW orders with momentum $\bm Q$ and hence breaks translational symmetry, a CDW order (with ordering wave vector $2{\bm Q}$) is naturally induced in the PDW state.\cite{Berg-2009,fradkin-2014} We propose that these Higgs modes can be observed in Raman spectroscopy, in momentum resolved electron energy loss spectroscopy, and in resonant inelastic X-ray scattering, which would provide evidence for the PDW order in the cuprates.

{We close with a few remarks on the role of disorder. In this paper we have only worked in the clean (disorder-free) limit. By including weak disorder,   momentum conservation is relaxed. Without the  constraints that momentum conservation implies for the kinematics, we expect that the gapless quasiparticle excitations in the pure PDW state or its coexistence state of $d$-wave SC to give rise to damping for both of the PDW Higgs modes. In particular, in the coexistence state, the density of states at the $d$-wave Dirac point becomes nonzero \cite{fradkin-1986a,fradkin-1986b,palee-1993} with disorder, which further enhances damping. On the other hand, disorder has much more serious effects on a translation symmetry breaking state such as the PDW superconductor. Indeed, in a layered material, such as the cuprate superconductors, the expected PDW state should be incommensurate. It is well known that in this case any amount of disorder destroys  true long range order,\cite{imry-1975} leaving behind a state with short range order and a vestigial (Ising) nematic orientational  order.\cite{nie-2013} The role of disorder in PDW superconductors and its consequences has so far been studied at a qualitative level (see  Refs.[\onlinecite{Berg-2009,senthil-mross}]). An interesting possibility is that the surviving vestigial order might be either nematic or a charge $4e$ superconductor \cite{Berg-2009b} (or both).
 The behavior of Higgs modes in conventional disordered superconductors has been studied in recent work.\cite{trivedi-2015}  An in-depth analysis of the disorder effects on PDW Higgs modes remains an interesting and important open question that we leave for future studies.}

\emph{Note}: After this work was completed we became aware of the recent paper by Boyack, Wu, Anderson and Levin\cite{Boyack-2017} who studied the collective mode contributions to the superfluid density in FF superconductors with a finite pairing momentum $\bf Q$. They found that, unlike in  uniform superconductors, the contribution from the Higgs mode in  FF superconductors is important, and that it destroys superfluidity well before the mean-field order parameter vanishes.

\begin{acknowledgments}
R.S.-G. and Y.W. contributed equally to this work.
We thank L. Radzihovsky for comments. 
This work was supported in part by FONDECYT (Chile) No.11160542 (RSG), by the U.S. Department of Energy, Office of Basic Energy Sciences under Award No. DE-SC0012368 (EF, SLC),
 and by the Gordon and Betty Moore Foundations EPiQS Initiative through Grant No. GBMF4305 (YW). \end{acknowledgments}

\appendix

\section{Alternative derivation for the Higgs mode energy in the FF state} 
\label{app:FF}

In this appendix we give the details for an alternative derivation of the Higgs mode energy in the FF state. We start from the polarization operator defined in Eq. \eqref{eq:bubbleFF}:
\begin{align}
K(\Omega_m)=&-\frac{1}{2}\int_{\bm k,\omega_m}\tr\[G({\bm k},\omega_m+\Omega_m)\tau_1G({\bm k},\omega_m)\tau_1\]\nonumber\\
=&\int_{\bm k}\frac{1}{\sqrt{4|\Delta_{\bm Q}^{(0)}|^2+(\xi_{\bm{k}}+\xi_{-\bm{k}+\bm Q})^2}}\nonumber\\
&\times\frac{(\xi_{\bm{k}}+\xi_{-\bm{k}+\bm Q})^2}{(\Omega_m^2+4|\Delta_{\bm Q}^{(0)}|^2+(\xi_{\bm{k}}+\xi_{-\bm{k}+\bm Q})^2)}
\label{a1}
\end{align}
and also define the polarization operator for the Goldstone mode (the phase mode):
\begin{align}
\bar{K}(\Omega_m)=&-\frac{1}{2}\int_{\bm k,\omega_m}\tr\[G({\bm k},\omega_m+\Omega_m)\tau_2G({\bm k},\omega_m)\tau_2\]\nonumber\\
=&\int_{\bm k}\frac{1}{\sqrt{4|\Delta_{\bm Q}^{(0)}|^2+(\xi_{\bm{k}}+\xi_{-\bm{k}+\bm Q})^2}}\nonumber\\
&\times\frac{(\xi_{\bm{k}}+\xi_{-\bm{k}+\bm Q})^2+4|\Delta_{\bm Q}^{(0)}|^2}{(\Omega_m^2+4|\Delta_{\bm Q}^{(0)}|^2+(\xi_{\bm{k}}+\xi_{-\bm{k}+\bm Q})^2)}.
\label{a2}
\end{align}
Note that the difference lies in the Nambu pseudospin structure inside the trace. 

Inside the FF state, the Lagrangian for the Goldstone mode can be obtained by performing an Hubbard-Stratonovich transformation on the four-fermion attractive interaction $V$ and integrating out fermions. Particularly, the
mass for the Goldstone mode is given by $1/V-\bar{K}(\Omega_m=0)$, which should vanish by definition. Therefore $1/V=\bar{K}(\Omega_m=0)$. On the other hand, the quadratic kernel for the Higgs mode Lagrangian, by the same procedure, is given by $1/V-{K}(\Omega_m)$, and setting $1/V-{K}(\Omega_m)=0$ gives the dispersion of the Higgs mode.

After the analytical continuation to real frequencies ($i\Omega_m\to\nu+i\eta$), we see from Eqs.\ (\ref{a1}) and (\ref{a2}) that the Higgs mode energy is given by
\begin{align}
 0&=\frac{1}{V}-K(\nu)=K(\nu)-\bar{K}(\nu=0)\nonumber\\
 &=\int_{\bm k}\frac{1}{\sqrt{4\Delta^2+(\xi_1+\xi_2)^2}}\left[\frac{(\xi_1+\xi_2)^2}{-\nu^2+4\Delta^2+(\xi_1+\xi_2)^2}-1\right]
\label{eq:higgsFF}
\end{align}
where we used the shorthand $\xi_1=\xi_{\bm{k}}$, $\xi_2=\xi_{-\bm{k}+\bm Q}$ and $\Delta=|\Delta_{\bm Q}^{(0)}|$. From Eq. \eqref{eq:higgsFF} we can read off that $\nu=2\Delta$, which is the same result in Eq. \eqref{eq:higgsFFenergy}.

\end{document}